\documentclass{article}
\usepackage[utf8]{inputenc}
\usepackage[T1]{fontenc} % Use 8-bit encoding that has 256 glyphs
\usepackage{fourier} % Use the Adobe Utopia font for the document - comment this line to return to the LaTeX default
\usepackage[english]{babel} % English language/hyphenation
\usepackage{amsmath,amsfonts,amsthm} % Math packages

\usepackage{array} % For better column definitions
\usepackage[table]{xcolor} % For cell coloring
\usepackage{adjustbox} % For adjusting box properties
\definecolor{lightgrey}{gray}{0.9} % Defines the light grey color

\usepackage{hyperref}

\usepackage[section]{placeins}
\usepackage[rightcaption]{sidecap}
\usepackage{graphicx} %package to manage images

\usepackage{xcolor}

\usepackage{sectsty} % Allows customizing section commands
\allsectionsfont{\centering \normalfont\scshape} % Make all sections centered, the default font and small caps
\usepackage{hhline}
\usepackage{multirow}
\usepackage{booktabs}
\usepackage{siunitx}

\title{Modeling Rational Adaptation of Visual Search to Hierarchical Structures}
\author{Saku Sourulahti, Christian P Janssen, Jussi PP Jokinen}
\date{\today}

\begin{document}

\maketitle

\begin{abstract}
Efficient attention deployment in visual search is limited by human visual memory, yet this limitation can be offset by exploiting the environment’s structure. 
This paper introduces a computational cognitive model that simulates how the human visual system uses visual hierarchies to prevent refixations in sequential attention deployment. 
The model adopts computational rationality, positing behaviors as adaptations to cognitive constraints and environmental structures. 
In contrast to earlier models that predict search performance for hierarchical information, our model does not include predefined assumptions about particular search strategies. 
Instead, our model's search strategy emerges as a result of adapting to the environment through reinforcement learning algorithms. 
In an experiment with human participants we test the model’s prediction that structured environments reduce visual search times compared to random tasks. 
Our model's predictions correspond well with human search performance across various set sizes for both structured and unstructured visual layouts.
Our work improves understanding of the adaptive nature of visual search in hierarchically structured environments and informs the design of optimized search spaces.

\end{abstract}

\section{Introduction}

Many daily tasks require visually locating pertinent information amidst potentially redundant, irrelevant, or distracting stimuli. For instance, finding a specific email in a crowded inbox, locating a product on a cluttered supermarket shelf, or identifying a key app icon on a smartphone's home screen are all tasks that demand effective visual search strategies. This is particularly true for various interactive technologies, such as computer desktops, mobile applications, and websites, whose displays are often filled with numerous visual elements.
Understanding how humans manage visual search tasks in complex search environments is crucial for designing such information spaces.
%A key question for establishing this understanding is: where should attention be deployed next to maximize information gain during search?
Humans are compelled to conduct visual search due to the information processing limitations of our visual system, which can only fixate on and encode a small, foveated subset of the full visual field at any given time \cite{findlay1998eye}.
To prioritize novel elements during search, the visual system employs a mechanism that marks and thereby inhibits returning to elements already searched and discarded, preventing redundant fixations \cite{posner1985inhibition,watson1997visual}.

Inhibition of return is supported by short-term visual working memory, which records previously visited elements and their features, aiding in directing attention to new elements \cite{jiang2004kind,mitchell2008flexible,vogel2001storage}.
This visual memory storage is limited, with the nature of memory capacity reported to be based on discrete units \cite{cowan2001magical, vogel2001storage, luck1997capacity} or a continuous pool of resources \cite{ma2014changing}, or the extension of internal memory to the external environment \cite{van2020embodied}.
However, the memory limitation poses a question: How is it possible to efficiently complete extensive visual search tasks when the number of distractors far exceeds the capacity of visual working memory? The ability to overcome short-term memory limitations during visual search can be partly attributed to the mind's capacity to form hierarchical representations \cite{baddeley2003working,ho2022people,hochstein2002view,oh2004role}.
Since such an ability seems possible due to the flexible nature of memory to adapt to structured environments, could this behavior emerge as a result of cognition's adaptation to the constraints imposed by the environment? 

\begin{figure}
    \centering \includegraphics[width=0.8\textwidth]{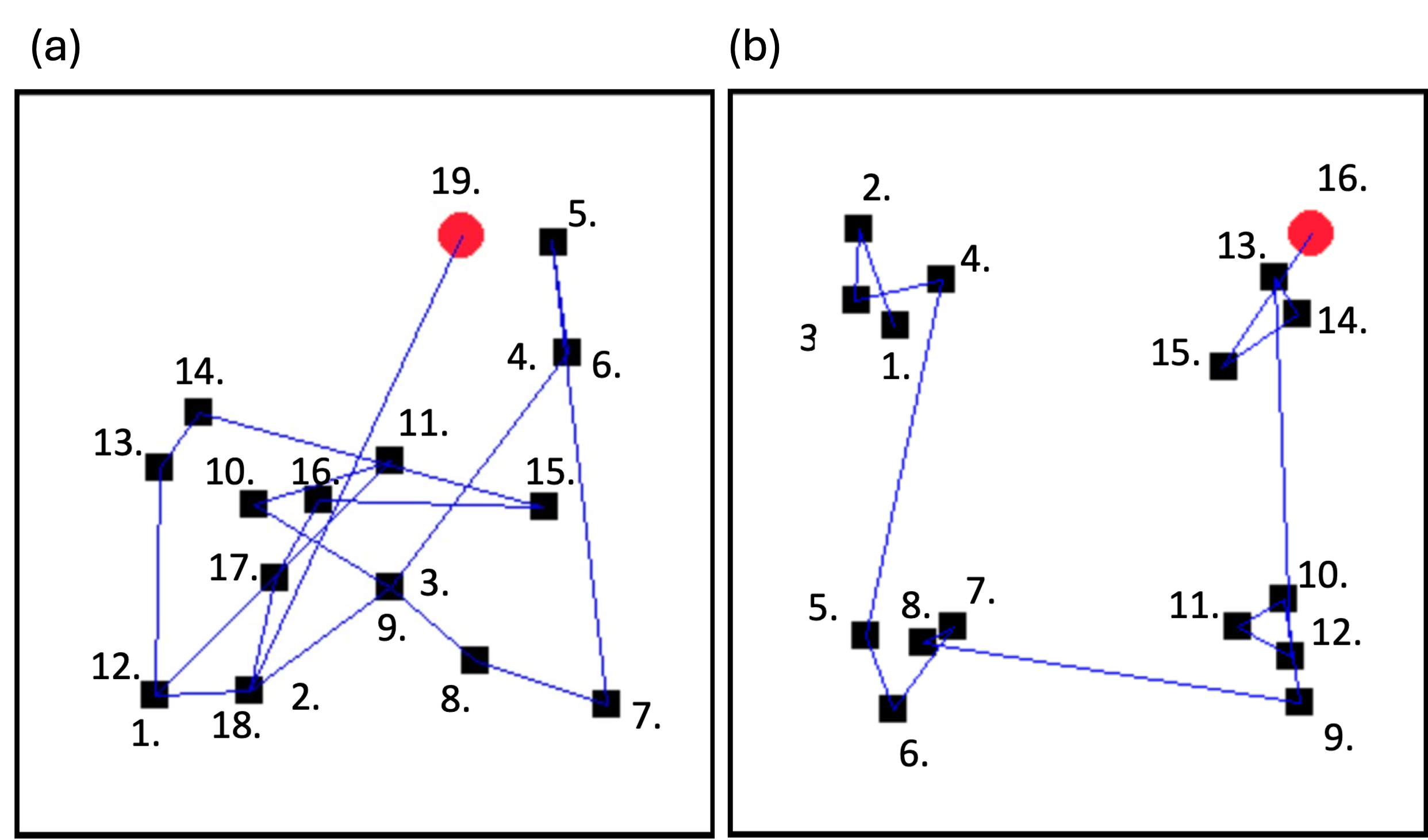}
    \caption{The model predicts the how eye movement trajectories adapt to visual task structure. Black rectangles are elements of the layout, and blue lines represent the eye movement search path from fixation to fixation in numerical order.  Left (a): search through randomly arranged visual elements. Right (b): Search through visually structured elements. 
}
\label{fig:path}
\end{figure}

This paper investigates and computationally models how the presence of perceivable visual structures impact visual search behavior. We propose that hierarchical visual structures simplify the task's representational complexity, enabling more effective utilization of limited visual memory. Our model can be seen in action in Figure \ref{fig:path}, which shows two stimuli side by side. In the first stimulus (a), visual elements are randomly arranged throughout the entire search area, resulting in model behavior where search is unstructured and previously marked distractors are revisited, thereby limiting efficiency. 
This is due to limitations in the model's visual working memory capacity. 
Conversely, in a task where the elements are perceptually grouped (b), model's search progresses first within-group, utilizing the limited visual working memory to exhaust the now smaller number of local elements, before moving to a new group of elements.
Due to a hierarchically structured visual working memory in the model, all elements of the previously searched group are marked by virtue of belonging to the same group.

A key aspect of our model is that the search behavior depicted in Figure \ref{fig:path} emerges naturally as a rational adaptation to the visual structure of the task and the limitations of visual working memory. Unlike predecessor models, our does not require explicit programming for search behavior \cite{teo2012cogtool, hornof2001visual}. This approach is grounded in computational rationality, which posits that behavior adapts rationally to cognitive constraints and the environment \cite{gershman2015computational,oulasvirta2022computational}.
In our framework, the search agent's internal cognitive environment possesses a hierarchical structure capable of retaining in short-term memory either individual elements or groups of elements.
The model dynamically adjusts to the hierarchical nature of its memory resources, resulting in a search strategy that prioritizes exhaustive exploration within a group before transitioning to other groups.

Prior computational modeling of visual search has covered a broad range of cognitive processes, such as bottom-up salience \cite{itti2000saliency, itti2001feature}, top-down expectations \cite{navalpakkam2005modeling, rao2002eye, sun2008computer, pomplun2003area}, and preattentive processes \cite{humphreys1993search, wolfe1994guided, deco2007attention}.
Our research complements these models by demonstrating the significance of visual hierarchies and how search behaviors adapt to them.
We make two contributions.
First, we present evidence that human visual search behavior rationally adapts to hierarchical visual structures within the task environment.
Second, our computational model is capable of simulating eye movement trajectories across visual search tasks, and as such offers a tool for \emph{in silico} evaluation of design candidates for various user interfaces.

\section{Related work}

\subsection{Hierarchical memory structure in visual search}

Extensive research in the field of visual attention has consistently demonstrated that the preattentive phase of grouping elements significantly enhances the efficiency of visual search processes. According to Feature-Integration Theory, information grouping occurs at a very early stage in the visual process, effectively separating relevant and irrelevant information from each other before the serial search process \cite{treisman1980feature, treisman1982perceptual}. 
This implies that visual groups are perceived as single units, thereby facilitating the memory of the group’s location, rather than requiring the separate retention of each item’s location in memory \cite{bundesen1983color, kahneman1992reviewing, pomerantz1973role}.

Visual memory is limited in remembering a fixed number of discrete items' colors and locations at the same time \cite{cowan2001magical, vogel2001storage, luck1997capacity}.
However, working memory has long been shown to enable an extended memory structure in which individual pieces of information are organized into larger units, a process referred to as visual chunking \cite{miller1956magical}.
This enables more efficient performance in tasks requiring visual memory, as task-redundant detailed information can be compressed into separate coherent units \cite{bays2009precision, brady2009compression}. Extended working memory, which integrates external single elements into memory, enhances perception by making it easier to recall grouped information as a single unit \cite{woodman2003perceptual}. Moreover, compressing visual information seems to be a rational strategy for improving perceptual performance to compensate for the limitations of visual short-term memory \cite{nassar2018chunking}. 
The extensive understanding of visual memory capacity and its flexibility to date  provides a clear picture of its ability to efficiently draw on information from the environment.

Computational models of working memory thus far would suggest that human attention seeks to operate efficiently at the global level of the visual perception, utilizing a hierarchical memory system \cite{brady2011hierarchical, brady2013probabilistic}. 
Working memory is more likely comprised of a multi-layered structural representation of visual perceptions rather than merely a collection of individual items \cite{brady2011hierarchical}, which is crucial for understanding the visual environment \cite{brady2013probabilistic}.
In visual perception, employing a hierarchical memory structure facilitates the rapid processing of information. This is supported by computational models and where subjects had to solve the traveling salesman problem from a display \cite{graham2000traveling, kong2010high}. 
These models achieve optimal performance even when visual objects are not organized into clear structures, as visual memory naturally tends to draw on structures from the environment \cite{graham2000traveling, kong2010high}. 
This suggests that the visual system systematically seeks the shortest eye-movement path, optimizing search both globally across groups and locally within each group, to make efficient use of the limited capacity of visual memory \cite{kong2010high}. 

Since it appears that cognition seeks to optimize visual performance through memory, it is evident that memory must retain elements that have already been encoded to reduce unnecessary revisits to the same elements. 
This phenomenon is also known as Inhibition of Return, which refers to the visual system's tendency to move to a new location to maximize information availability and optimize the search path with limited memory capacity \cite{posner1984components, posner1985inhibition}.
The findings indicate that the visual system is capable of inhibiting 4-5 items simultaneously \cite{snyder2000inhibition, wang2010searching, cowan2001magical}, which might suggest a limited capacity of memory. 
In any case, it appears that the IOR phenomenon has a significant impact on facilitating the discovery of the shortest possible visual search path \cite{klein1999inhibition}.

Given the importance of inhibition of return in optimizing visual search, our main hypothesis is that the visual system strives to use memory efficiently to minimize the encoding of items as effectively as possible. 
Instead of attempting to remember the locations of all individual visual items with limited memory capacity, it's possible to retain only the locations of visual areas in memory. 
In cognitive research on visual attention, it has not yet been demonstrated how the cognition adapts to leverage visual structures when the goal is to achieve the optimal search performance. 
Furthermore, research field has not yet been able to demonstrate how, as a result of adaptations, the increase in IOR emerges through hierarchical memory to maximize encoded areas, especially when the search task emphasizes visual structures.

\subsection{Models of Visual Search}

Computational modeling plays a significant role in understanding visual search, offering insights into the cognitive processes of attention by predicting eye movement behavior.
Visual search models vary according to the task goal and visual stimuli, with their predictions usually relying on the integration of bottom-up and top-down processes.
Salience map models offers an explanatory framework for the bottom-up process by directing immediate attention to visually salient features that stand out [15, 16].
Salience information based on the phenomenon that stimulus-based visual features initially attract attention, with significant visual elements then competing for gaze attention.
Another crucial aspect of the modeled attentive process is the top-down approach, in which a 'target map' assigns probabilities to features that are visually or semantically similar to the target \cite{navalpakkam2005modeling, rao2002eye, sun2008computer, pomplun2003area}.
Target map-based models guide eye movement decisions towards more likely target locations. This process involves the goal-directed aspect of attention, guided by the individual’s current goals, expectations, or knowledge. 

Although models offer a relatively successful predictive framework for major cognitive attention processes, it is crucial to recognize that the nature of the search task requires an emphasis beyond merely top-down and bottom-up processes.
Most existing search models fail to adequately consider the importance of visual structures in visual search, as targets in traditional 2D tasks are almost immediately visible \cite{li2018memory}.
Although such computational models account for the memory of the most recent fixation locations \cite{itti2001feature, parkhurst2002modeling, navalpakkam2005modeling}, only a few models emphasize the advantages of structured information for improving Inhibition of Return.

To date, only a few hierarchical visual search models have been created to replicate how semantic structures influence to shorter search paths \cite{teo2012cogtool, hornof2001visual}.
These models are grounded in cognitive architectures such as ACT-R, EPIC, and ACT-R/PM, and rely on production rules to guide eye movement decision-making through determined logic.
In the majority of models, the memory of the recent locations is directly integrated into the decision-making process through production rules or by reducing the probabilities for the next fixation location. In particular, the visual search of hierarchical models mimics a certain search strategy, rather than the model itself adapting to the optimal search strategy in certain the task environment.
The explanatory power of the adaptive search model illustrates how, due to cognitive constraints, the inhibition of encoded elements emerges, as seen in the Adaptive Feature Guidance Model \cite{jokinen2020adaptive}. However, existing search models have not demonstrated how the inhibition of scanned locations is enhanced through rational adaptation to visual structures, assuming that visual memory structure is hierarchical.

Our model's cognitive approach is based on the Computational Rationality, wherein the model learns to optimize the search path of eye movements by adapting to the task environment in accordance with cognitive constraints and capacity \cite{gershman2015computational}. Computational rationality makes a substantial difference to almost all of the above models predicting hierarchical search, from which the decision to shift attention arises.
In models grounded in cognitive architectures, this means that production rules directly determine the constraints for the decision-making of the subsequent fixation.
In map-oriented models, an updated probability map (salience/target map) directs eye movement based on salience and target features.
In such models, the decision-making for eye movement draws on statistical mathematics, machine learning algorithms, or reinforcement learning.
It is noteworthy that decision making in our model emerges through reinforcement learning having available environmental information.
Internal representation from environment includes the egocentric information on the distances of fixation to the elements, grouping of elements, and a hierarchical representation of the locations of visited elements. 

In our study, we assume that human perceptual ability can form coherent structures even from randomly placed elements \cite{brady2013probabilistic}.
Contrary to most visual search models, our model does not bring out the top-down process expectations for target features or the bottom-up process's salience features to guide attention because these are not essential processes for the task environment we are interested in.
Our model aims to maximize the efficiency of the search task by utilizing a hierarchical structure of visual memory with a global search strategy for spatially grouped structures.
This is achieved through a hierarchical representation that stores entire grouped structures rather than the locations of individual elements independently.
Thus, the phenomenon of Inhibition of Return arises as a result of the model adapting to environmental structures, rather than, as in most models, by directly implementing prevention of revisits into the decision-making logic or by reducing probabilities of encoded locations.  

\subsection{Goals of the paper}

In this study, we identify three distinct main objectives: 1) To demonstrate through empirical observations that utilizing visual structures improves performance in search tasks, 2) To develop a computational model that replicates human behavior in structured layouts, 3) To demonstrate the model's potential as a tool in HCI design to aid designers' decision-making. 

The aim of the study is to demonstrate that human behavior aligns with our assumptions about cognition adapting to environmental visual structures using a hierarchical memory structure, to gain an advantage in visual search performance. 
Empirical findings aim to prove how visual structures contribute to faster target detection.
Above all, we show with our model that the assumed hierarchical a memory structure facilitates a more efficient search strategy by enhancing the inhibition of previously observed elements.
The model's prediction enables future applications to assist UI designers in optimizing information categories into visually appropriate group sizes by simulating the user's eye movement for a given spatially structured layout.

\section{Model description}

\subsection{Computational rationality and reinforcement learning}

A visual search involves strategic planning to optimize search performance with respect to the task's goal \cite{najemnik2005optimal}. 
The fundamental basis of the model is that the agent independently learns the optimal course of action through goals, possible actions, and observed environmental states. 
In this way, the agent's decision-making becomes strategic, not settling for the immediately shortest next fixation but instead looking further into the future. 
The decision-making can be considered 'rational,' as the central idea of the model is the agent's ability to rationally predict the most favorable successive actions, aiming to achieve the maximum benefit from the task environment \cite{gershman2015computational}. 
However, with limited internal representation, the benefit can only be achieved by exploiting the available information as much as possible, but at the cost of computational complexity \cite{gershman2015computational}. 
The agent's decision-making is constrained by the limited knowledge of environmental states, possible actions available to the agent, the transition function between possible states and the reward function. 
Computationally rational learning is determined through reinforcement learning algorithms, whereby the agent interacts with the environment making rational choices based on the available information and its own computational limitations, aiming to achieve the maximum reward for the task \cite{oulasvirta2022computational}. 

The model's decision-making process operates as a Partially Observable Markov decision Process (POMDP), with the agent's decision-making guided by an internal representation. 
POMDPs is an approach for modeling sequential decision-making processes, where achieving a goal requires several consecutive steps where each step depends on previous steps \cite{oulasvirta2022computational}. 
The POMDP decision process observes the environment partially, limited by human cognitive constraints in obtaining information from the environment. 
POMDP describes a sequential stochastic decision process, where the tuple <S, A, T, R, \(\gamma\)> consists of a finite set of states (S), a finite set of actions (A), a transition function (T), a reward function (R), and a discount factor ($\gamma$).
The process controls decision-making by selecting the most favorable action for the long-term goal from all possible actions $a \in A$ in a certain state, through interaction with the environment. 
The transition function defines the probability T(s, a, s\(\prime\)) = p(s | s, a) for the future state \( s^\prime \in S \) upon selecting an action in a given current state $s \in S$. 
A "policy" is a current strategy or a set of rules that an agent follows to determine its actions based on the current state of the environment. 
The agent attempts to optimize its policy \(\pi\) by maximizing the long-term reward and selects actions by following the current policy, which defines the probabilities for actions in a given state as \(\pi\)(s, a) = p(a | s). 
The reward function defines the probability R(s, a) = p(r | s, a) of achieving a reward $r \in R$ with a chosen action, once the action has been executed.

The agent's policy evolves as its performance optimizes, and in accordance with the Bellman equation, the maximum expected return (including immediate and discounted future rewards) that can be achieved from state s onward. 
This is achieved by optimal policy which maximize the value function:  

\[
V^*(s) = \max_a \left[ R(s, a) + \gamma \sum_{s' \in S} T(s, a, s') V^*(s') \right]
\]

where, \(\gamma \in [0, 1]\) denotes the immediate rewards are discounted in relation to the future. Since the model adheres to an optimal strategy for future actions, it strives to select actions that maximize the sum of future rewards. 
The agent is tasked with searching for a randomly selected target among the visual elements. 
The choice of the next fixation is made from available actions \(a \in A\), which corresponds to the locations of all elements.

Model's rational learning is determined by aforementioned theoretical description of reinforcement learning. 
Computation rationality entails the architectural structure of the model, where the agent's internal environment interacts with the external environment. Agent’s rational choices based on available information and its own computational limitations, aiming to achieve the maximum reward for the task at hand \cite{ oulasvirta2022computational}. 
In the following section, we will walk through how interaction occurs within the model's architecture.

\subsection{Architecture of model}
\begin{figure}
\centerline{\includegraphics[width=\linewidth]{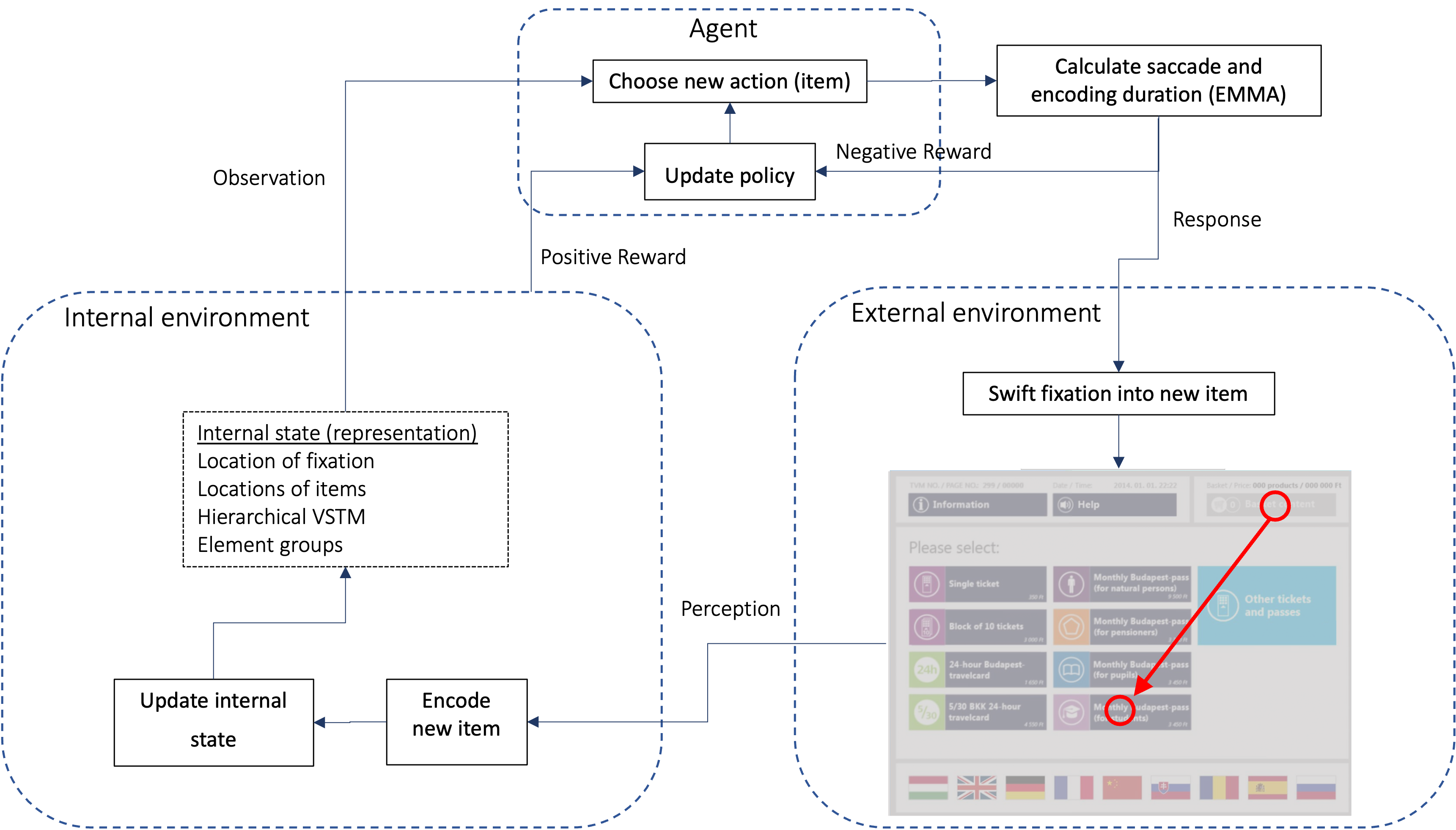}} % Adjust the file extension as per your image type
\caption{The model's architecture, where the agent makes decisions based on the state of the external environment, but through the internal environment's state. 
This internal state changes via the perception of stimuli into an internal representation relative to cognitive capacity and constraints. 
The agent's new action shifts the fixation to the location of a new element, and the duration taken for the saccade is calculated using the EMMA model, which results in a negative reward for the agent. 
A new element, in response, causes a new state change in the external environment.\label{fig:fig_model_arc}}
\end{figure}

Our model's architectural structure can be roughly divided into agent, internal and external environments (Figure~\ref{fig:fig_model_arc}). 
The agent is in constant interaction with the external environment indirectly through an internal cognitive representation, which is agent’s belief state. 
The internal environment encompasses a cognitive space that provides a partial visual representation of the external environment \cite{oulasvirta2022computational}. 
The external environment refers to the physical context of the interaction, which in this instance mean the spatially structured layout. 
The agent implicitly utilizes the information from the internal environment's state, making the changes in states stochastic. 
These changes depend on the agent's probabilities of selecting a certain action in a given internal state. 
The agent observes and evaluates changes in the internal environment's states, receiving positive and negative rewards. 
The agent's action cause response to the external environment, leading to a change in its state, which modifies the representation of the internal state through the perception. 
Next, we will delve more deeply into the internal and external environment’s mutual interaction.

The agent makes observations from the internal environment and decides which element to move attention to next. 
This process requires information from the internal environment to enable the optimization of its search policies. 
Since the model's internal visual representation of the environment is partial, it means that percepts of the external environment are constructed with cognitively limited capacity from stimuli \cite{oulasvirta2022computational}. 
Internal representation is predefined in terms of the actual external environment, making the model's input information symbolic. 
This implementation is because the model aims to demonstrate its adaptation to the environment, where it learns to utilize visual information for performing search tasks rather than focusing on how the representation is formed through low-level perceptual processes from raw pixel data.

The model's internal representation is not directly available for the agent's decision-making but is used implicitly. 
Figure~\ref{fig:fig_model_arc} illustrates the cognitive features included in the internal representation. 
The model adapts to the environment by utilizing information available to the agent about 1) location of current fixation ('Eye'), 2) the locations of elements (‘Element distances’), 3) leveraging the hierarchical visual short-term memory (‘Hierarchical VSTM’), and 4) the grouping of elements (‘Groups of elements’). 
The model directs its attention to one element at a time, and the attention’s information in the representation is updated each time the agent takes a new action towards a new element's location. 
The egocentric representation of elements' locations enables the model to search for a shorter path. 
Using this information, the model learns to plan the length of the eye path, aiming to cover as many elements as possible within a given time.
Spatial relationships in internal representation have been simplified to distance information between current fixation and other elements. 
This reduction in information is sufficient to allow the model's computational capacity to be adapted.

The agent learns to optimize the length of the search path by doing enough consecutive observations of the environment and improving the selection of action in each state.  
To strategically solve the optimization problem for the long term rather than just moving to the nearest next element, the agent must be implicitly provided with information about encoded element locations in short-term memory. 
This information ultimately guides decision-making to inhibit already scanned elements that would otherwise lead to unnecessary revisitation of the same elements. 
However, merely simple short-term memory is not sufficient for efficiently performing search tasks on layouts with larger numbers of elements. 
Therefore, the model leverages a Hierarchical VSTM to enhance the efficient use of memory by maximizing the memorization of scanned element locations. 
This is possible for model by the element grouping information of elements, which identifies which elements belong to a same group. 
When all the elements of a group are encoded within a given visual group, the location information of the entire group can be encoded into the higher hierarchical memory level. 
Once the agent has recognized the advantage of grouping, it ultimately seeks to plan the search path primarily as a global search strategy.

The internal environment defines the reward received by the agent relative to the goal of the task, which in this instance is finding the target as fast as possible. 
The model's agent receives negative rewards according to the time spent on eye movements. 
The agent's learning emerges when it has encountered the mentioned associations enough times and achieving sufficient positive reward from the internal representation. 
The agent's adaptation for certain strategy is dependent on how much greater a reward the new policy offers, such as leveraging hierarchical memory over leaning solely on the limited capacity of short-term memory.

The duration of the model's eye movement shift is estimated using the EMMA eye movement model \cite{salvucci2001integrated}, which determines the time spent on saccades and encoding. 
Since EMMA does not account for encoding durations for different types of symbols, we added a 150ms duration to each fixation to fit the human data. 
This is justified, as fixation time typically ranges from 200-250ms when reading, and overall fixation duration can vary between 100-500ms \cite{rayner1978eye}. 
Given that the average word length is longer compared to our 3-letter combinations, it can be argued that fixation would be significantly shorter than 200ms.

\subsection{Internal representation}

\begin{figure}
\centerline{\includegraphics[width=0.85\textwidth]{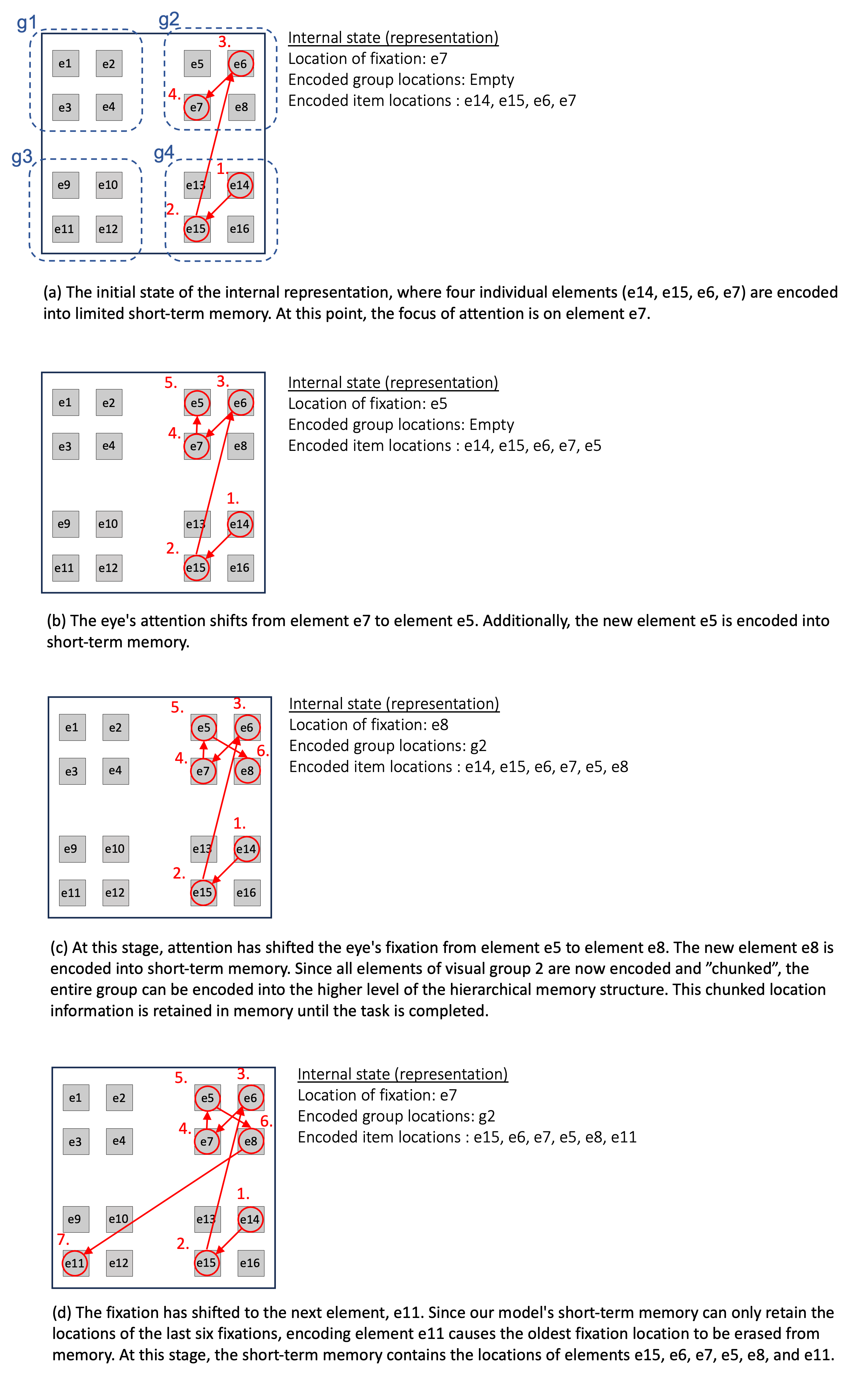}}
\caption{The series of figures illustrates the change in the model's internal state as the simulated eye movement proceeds from element to element, aiming to encode one spatial group at a time. 
Eye-movement figures demonstrates how our model's encoding occurs when it is optimized to use hierarchical VSTM optimally. 
The model’s internal state also includes static information about visual grouping and the spatial locations. 
The grouping of elements is bounded by a blue dashed lines and group numbers are presented in the first stage (a) of the series of images.\label{fig:fig_state}}
\end{figure}

This section presents an example of the model's simulation (Figure~\ref{fig:fig_state}), illustrating how the model's optimized policy is reflected in changes to the internal representation state. 
In this simulation example, the model has learned to utilize visual structures and is therefore able to efficiently make use of the hierarchical structure of visual memory. 
Our model's internal representation includes four observation spaces: 1) Eye position 2) Egocentric distances to elements 3) Hierarchical memory structure of element locations 4) Grouping of elements. 
To simplify the figure~\ref{fig:fig_state}, we omitted the grouping information and egocentric spatial data from the internal state. 
The model learns a policy that aims to plan the search path through one group at a time.
This grouping information remain as static information throughout the task, as the grouping details emerge during the preattentive stage of the bottom-up process in the peripheral visual field \cite{treisman1980feature, treisman1982perceptual}.  

From the first fixation (a), the eye information in the representation is updated to the element for which the action is selected according to the policy. 
At the same time, each new fixation is encoded into short-term memory. 
In stage (c), the model has encoded all the elements of an entire group into short-term memory. 
As a result, the group-specific information is encoded at a higher level of the hierarchical memory structure, which remains in memory until the end of the task. 
In this case, we assume that memory chunks are more likely to remain in long-term memory, and that one of the advantages of chunks is assumed to be their easier retrieval from memory compared to retrieving individual elements from short-term memory \cite{miller1956magical, chase1973mind, nikolic2007creation}. 
In the last stage d, after the seventh fixation, the oldest encoded element is erased from short-term memory, and the new fixation or element location is added. 
This is because VSTM can retain only a limited number of 6 encoded elements.

\section{Evaluation}

\subsection{Method}
In the experimental design, all subjects performed all the visual search tasks (within subjects).
In the experiment, participants performed visual search tasks where they had to find a target letter combination that appeared on the screen as quickly as possible from the stimulus and press the reaction button as soon as they found it. 
The experiment consisted of 6 different tasks, and each task included stimuli for four different element sizes (16, 24, 36, 48), varied for both structured and unstructured conditions (a total of 48 tasks). 
The stimulus for each search task were also unique, generated by a computer program.

\subsection{Participants}
The study was online, which enabled a huge number of participants (48 subjects) aged between 18 and 29 years. All participants were either U.S. or U.K. citizens, ensuring fluent English language skills for understanding the instructions. 
During recruitment, it was ensured that all participants had good vision and no abnormal disabilities or impairments that could affect their physical reaction abilities. 

\subsection{Procedure}
Before the actual tasks began, participants were asked to first accept the consent form. 
In addition, the participant went through the instructions for the application task, which explain how each task should be perform. 
Participants were also instructed to eliminate all extraneous distractions to focus solely on the experiment. 
They had the option to terminate their participation in the study at any point during the experiment. 
At the beginning of each individual visual search task, a 4-second countdown timer appeared on the screen, signaling participants to prepare by placing their finger on the space bar (reaction button). 
After that, a letter combination (target) appeared on the screen for 3 seconds, which participants were asked to memorize. 
The target letter combination was randomly selected from a set of letter combinations in the layout for each task. 
Following this, the actual stimulus appeared on the screen, and the participant had to press the spacebar as quickly as possible upon finding the target letter combination. 
This was immediately followed by a new search task, which again started with a 4-second countdown. 
The experiment included one practice task and a total of six measurable tasks. 
Each task included search tasks for all element set sizes (16, 24, 36, and 48) combined with structured and unstructured layouts. 
The order of the tasks was Latin-square counterbalanced, and the layout of each individual search task was unique. 
There was an opportunity to take a 30-second break between each task. 
Overall, the experiment took about 15 minutes.

\subsection{Stimuli}
The generation of stimuli for both the experiment and the model has been carried out by a computer program. The layouts used in our study contain letter combinations of the same color (including the target string), which represent visual elements. 
The elements are either in perceptible, distinct spatial groups (structured) or in non-distinguishable groups, meaning that the elements are evenly distributed across the entire layout area. 
We assumes that human perceptual ability can form coherent structures even from randomly placed elements \cite{brady2013probabilistic}. 
This means that in our model, the groups are less discernible in the unstructured layout than they would be in a structured layout. 
In the internal representation, this means that elements belong to only one group in a structured layout, but in an unstructured layout, elements can belong to more than one spatial group. 
In the Appendix A (The generation of the layout), it is described in detail how the structures of the layouts are generated and how the spatial groups of the model's external environment are mapped to the model’s internal representation.

\subsection{Results}

\subsubsection{Experiment}
In terms of the results, we were interested in how the number of elements in a layout affects human search time and how this varies when visual information is either structured or unstructured. 
Secondly, we were interested in how the search time of a model performing a search task aligns with the human search time for the same task under the same stimuli. 
The initial assumption about how the spatial structures of elements affect search performance was based on two hypotheses: 1) an increase in the number of elements in the layout increases search time, and 2) an unstructured layout increases search time relative to a structured layout for a given set size. 
Next, we will walk through how well these hypotheses were realized based on the results.

\begin{figure}
\centerline{\includegraphics[width=0.9\textwidth]{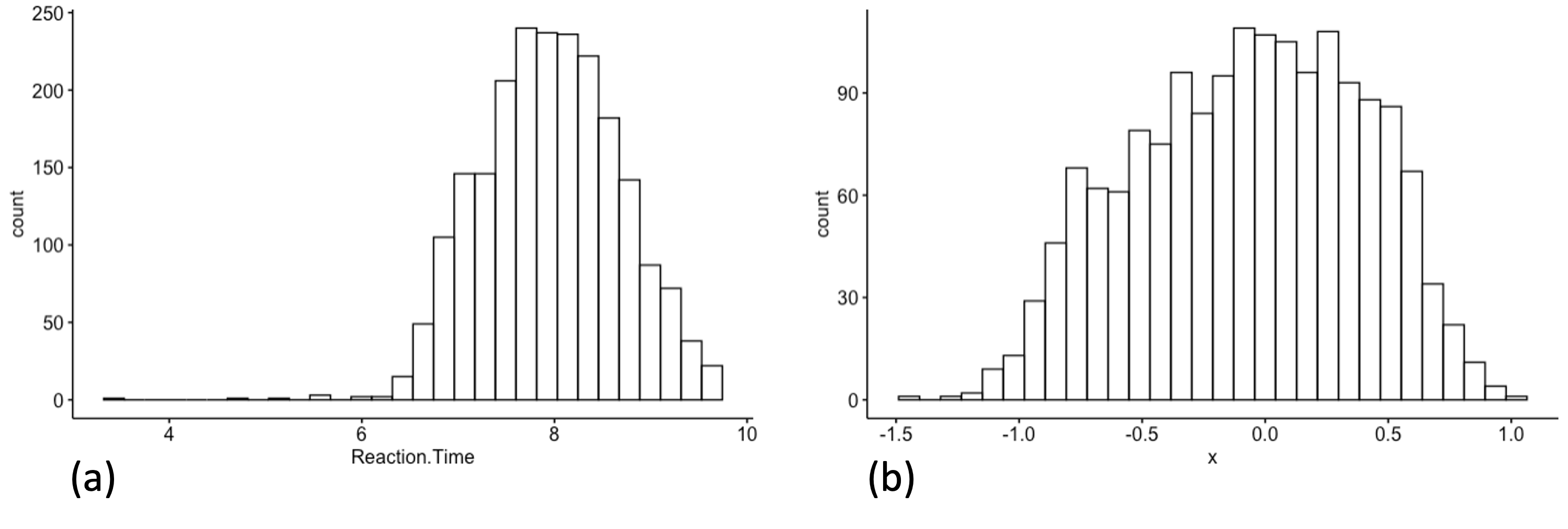}}
\caption{Logarithmically transformed distribution of the experiment results on the left (a). Logarithmically transformed distribution of residuals from the linear mixed model shown on the right (b). \label{fig:log}}
\end{figure}

A total of 2304 trials were collected from 48 subjects in the study. 
We used the Interquartile Range (IQR) filtering method for outlier removal, after which a total of 2155 trials remained for the statistical analysis of the data. 
In total, 149 outliers were removed from the data, which is about 6.4 percent of all the original trials. 
The original distribution of the experimental data indicated that there was a clear skewness in the search time distribution, which was transformed logarithmically to appear normally distributed (Figure \ref{fig:log}.a) \cite{osborne2010improving}. 

In analysis, we want to account for random variation between participants, so we conducted a statistical analysis using a linear mixed model. 
This model allows for the evaluation of both fixed and random effects on search time. 
Through the statistics model, we aim to determine how independent variables (set size and spatial organization) and their interactions affect the variation in search time. 
Set size and condition are fixed effects of the model, and the variation between participants is a random effect to explain the model's search time.

The random effect of the linear mixed model does not seem to cause a large standard deviation in the impact of individual differences (0.57 s). 
The effect of individual differences on the random variance is a very small fraction of the unexplained variance in the overall model (ICC=0.05). 
The model partially explains the variation in results in the experiment. 
Although the residual variance value (5485.02 s²) and the residual standard deviation (2.34s) are fairly large, the distribution of residuals appears fairly normally distributed when the data is logarithmically transformed (Figure \ref{fig:log}.b).

\begin{figure}[b!]
\centerline{\includegraphics[width=0.6\textwidth]{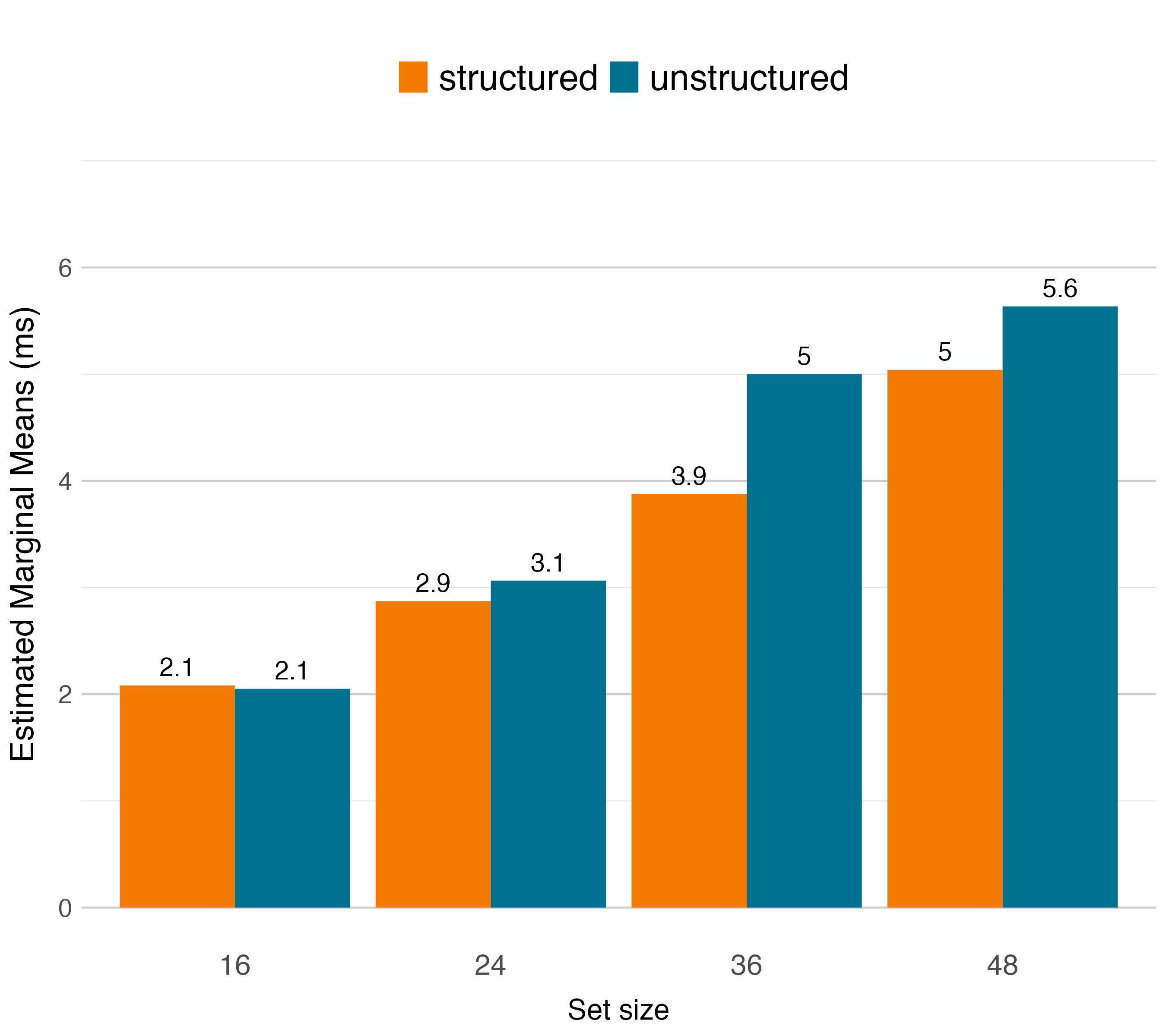}}
\caption{Estimated marginal means search times for each conditions.\label{fig:emmans}}
\end{figure}

Firstly, figure \ref{fig:emmans} shows that search time increases as expected with the growth of set size \cite{treisman1980feature, wolfe1994visual}. 
ANOVA analysis of the linear mixed model revealed that set size has a significant increasing effect on search time, with larger set sizes leading to longer search times (F(3,2101.4)=205.72, p<0.001). 
Secondly, the results indicate that a structured organization facilitates finding the target more quickly. 
A structured organization also has a significant effect on improving search performance (F(1,2101)=21.25, p<0.001). 
However, the advantage of a structured organization in search performance appears to improve more only with a sufficiently large set size, and it becomes more pronounced with even larger set sizes. 
This interaction effect between set size and spatial organization is significant (F(3,2101)=6.16, p<0.001). 
In Figure \ref{fig:emmans}'s estimates, it can be observed that with the smallest set size (16), a structured organization does not provide any benefit in search performance, as the search time was 2.1 seconds in both conditions. 
With a set size of 36, the search time for the structured condition (m = 3.9s) is already significantly better compared to the unstructured condition (m = 5.0s).

\subsubsection{Model Evaluation}
The computational model's predictions align well with human search performance in tasks involving visual structures. 
We analyzed the computational model’s ability to predict human search times using a linear regression model. 
In this analysis, the search time produced by the model simulation is the independent variable, while participant’s search time is the dependent variable. 
The predictions of the regression model show that the relationship between the computational model and human search times is strong. 
According to the results, the computational model's ability to predict participants' reaction times was significant (t(6) = 8.064, p < 0.001). 
When the model's search time increases by one second, human search time increases by 0.43 seconds. The model's intercept was also statistically significant (t(6) = 5.591, p = 0.0014).
The analysis revealed a high adjusted R-squared (R² = 0.90), indicating that the predictions of our computational model explain about 90\% of the variation in human search times. 
The prediction of the linear model is statistically highly significant (F(1, 6) = 65.02, p < 0.001), indicating strong predictive power of the computational model with respect to human performance in the given task. 
Although the model's predictions are generally accurate for each condition, there is a slight mean error in the linear model's predictions compared to the actual search times (RMSE = 0.38 seconds).

\begin{figure}
\centerline{\includegraphics[width=0.8\textwidth]{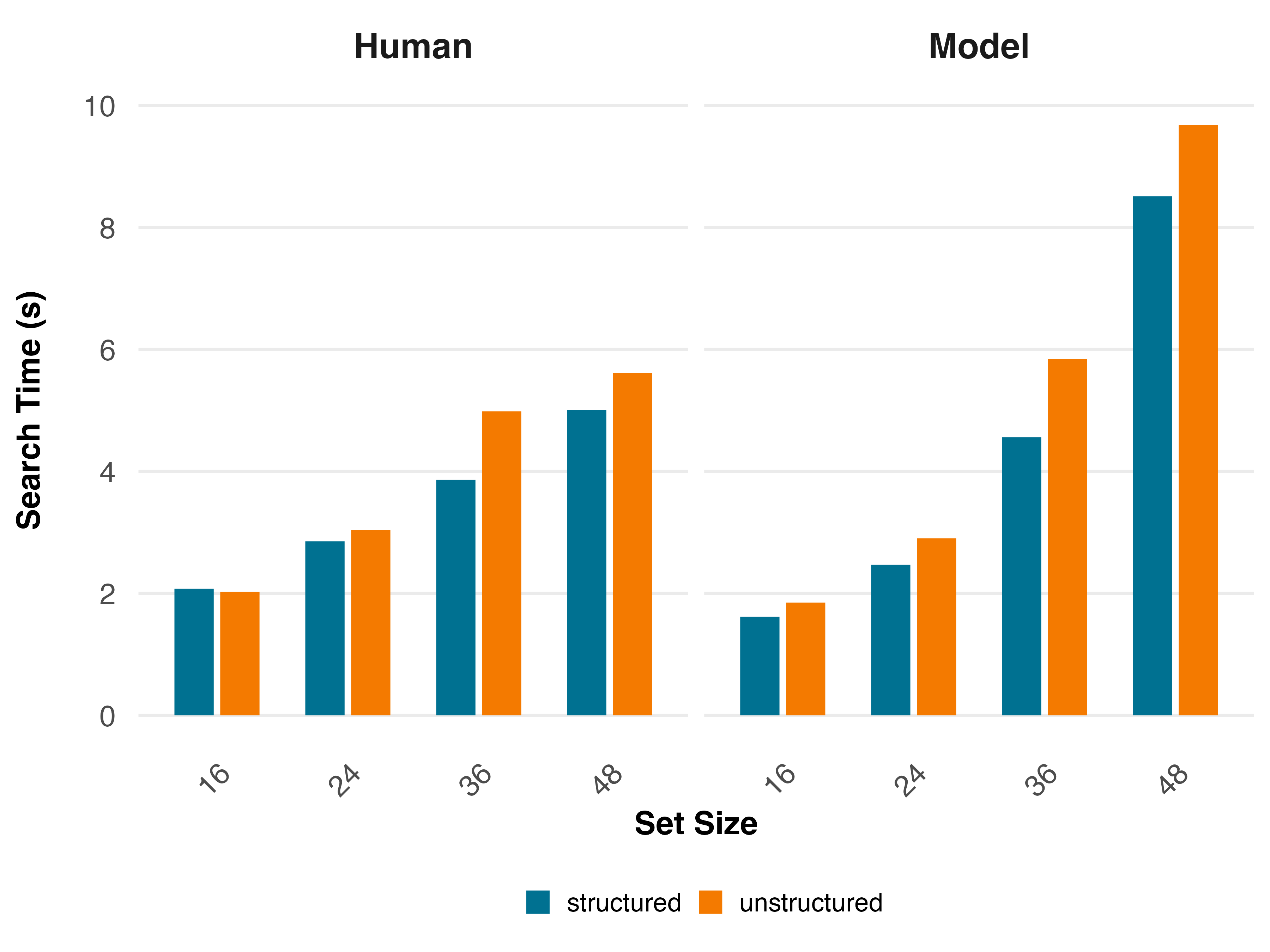}}
\caption{Mean search times for human and model data across set sizes under structured and unstructured conditions.\label{fig:hum_mod_ave}}
\end{figure}

In the comparison of the model and experimental data (Figure \ref{fig:hum_mod_ave}), it can be observed that the means well support our computational model's ability to predict the advantage of a structured layout in search performance and the effect of a larger set size on the increase in search time. 
The improvement in search time in the structured condition becomes increasingly apparent as set size increases, both for the model and for humans. 
This suggests that the model effectively captures the interaction between set size and the structured condition in influencing search performance.

Due to the model's adaptability, it is capable of accounting for the stochastic nature of target detection. 
It is more likely that the target element will be found faster even randomly with a smaller set size, in which case structures do not have enough time to benefit memory in inhibiting items. 
This can be observed with a sample size of 16, where search times are almost at the same level in both the structured and unstructured layout for both the computational model and the experiment. 
With larger set sizes, the differences between the structured and unstructured layouts become more pronounced in both the model and the experiment. 

However, the model's search time increases significantly with larger set sizes, due to its inability to learn the utilize of structures in larger set sizes. 
In other words, the model’s computational limits for adaptation are attained with larger set sizes, although these challenges can be solved in the future. 
Another significant factor contributing to the model's excessively long search times in larger set sizes is the limited two-level hierarchy of its visual memory. 
In the model, this phenomenon is emphasized with larger set sizes, as the visual memory can only form chunks of limited size. These points are discussed in more detail in the Discussion section.

The results of the simulations seem to provide support for our underlying assumption that cognition adapts to use a hierarchical memory structure, as it more efficiently improves memory recall of previously encoded elements and avoids revisiting to elements. 
Firstly, this phenomenon is supported by the results in Table \ref{tab1} on simulated fixations, which show that fixations decrease due to the effect of a structured organization for a given set size. 
The number of fixations increases the difference between the structured and unstructured conditions for larger set sizes, which appears to be well aligned with the results of the model on search performance.

Secondly, the increase in IOR predicted by the effect of structured organization supports a smaller number of revisits. The difference in revisits becomes more apparent only with larger set sizes (36 and 48). 
Since the differences in fixations and revisits between structured and unstructured layouts are still small, a slightly larger increase in fixations does not necessarily lead directly to a significantly greater number of revisits. 
Also, as expected, an increase in set size leads to an increase in the number of revisits, as finding the target naturally takes longer.

\begin{table}[ht]
\centering
\caption{Summary of fixations and revisits across structured and unstructured conditions}
\label{tab1}
\begin{tabular}{lllll}
\toprule
\textbf{Spatial Organization} & \textbf{Set Size} & \textbf{Fixations} & \textbf{Revisits} & \textbf{Revisits SD} \\
\midrule
Structured   & 16  & 8.6  & 0.2  & 0.8   \\
Structured   & 24  & 12.7 & 1.1  & 2.3   \\
Structured   & 36  & 21.9 & 3.5  & 5.4   \\
Structured   & 48  & 36.9 & 12.3 & 17.1  \\
Unstructured & 16  & 8.8  & 0.3  & 1.0   \\
Unstructured & 24  & 12.9 & 1.0  & 2.7   \\
Unstructured & 36  & 27.9 & 9.7  & 14.1  \\
Unstructured & 48  & 38.5 & 16.2 & 23.1  \\
\bottomrule
\end{tabular}
\end{table}

%\begin{center}
%\begin{table}
%\caption{Summary Statistics of Revisits and Fixations.\label{tab1}}
%\begin{tabular*}{\textwidth}{@{\extracolsep\fill}lllll@{}}
%\toprule
%\textbf{Spatial Organization} & \textbf{Set Size}  & %\textbf{Fixations}  & \textbf{Revisits}  & \textbf{Revisits SD}   \\
%\midrule
%Structured   & 16  & 8.6  & 0.2  & 0.8   \\
%Structured   & 24  & 12.7 & 1.1  & 2.3   \\
%Structured   & 36  & 21.9 & 3.5  & 5.4   \\
%Structured   & 48  & 36.9 & 12.3 & 17.1  \\
%Unstructured & 16  & 8.8  & 0.3  & 1.0   \\
%Unstructured & 24  & 12.9 & 1.0  & 2.7   \\
%Unstructured & 36  & 27.9 & 9.7  & 14.1  \\
%Unstructured & 48  & 38.5 & 16.2 & 23.1  \\
%\bottomrule
%\end{tabular*}
%\begin{tablenotes}%%[341pt]
%\item[$^{\rm a}$] Summary of fixations and revisits across structured and unstructured conditions.
%\item[$^{\rm b}$] Set sizes of 16, 24, 36, and 48 are used for each condition.
%\item {\it Source}: Derived from experimental data.
%\end{tablenotes}
%\end{table}
%\end{center}

\begin{figure*}[b]
\centerline{\includegraphics[width=1.0\textwidth]{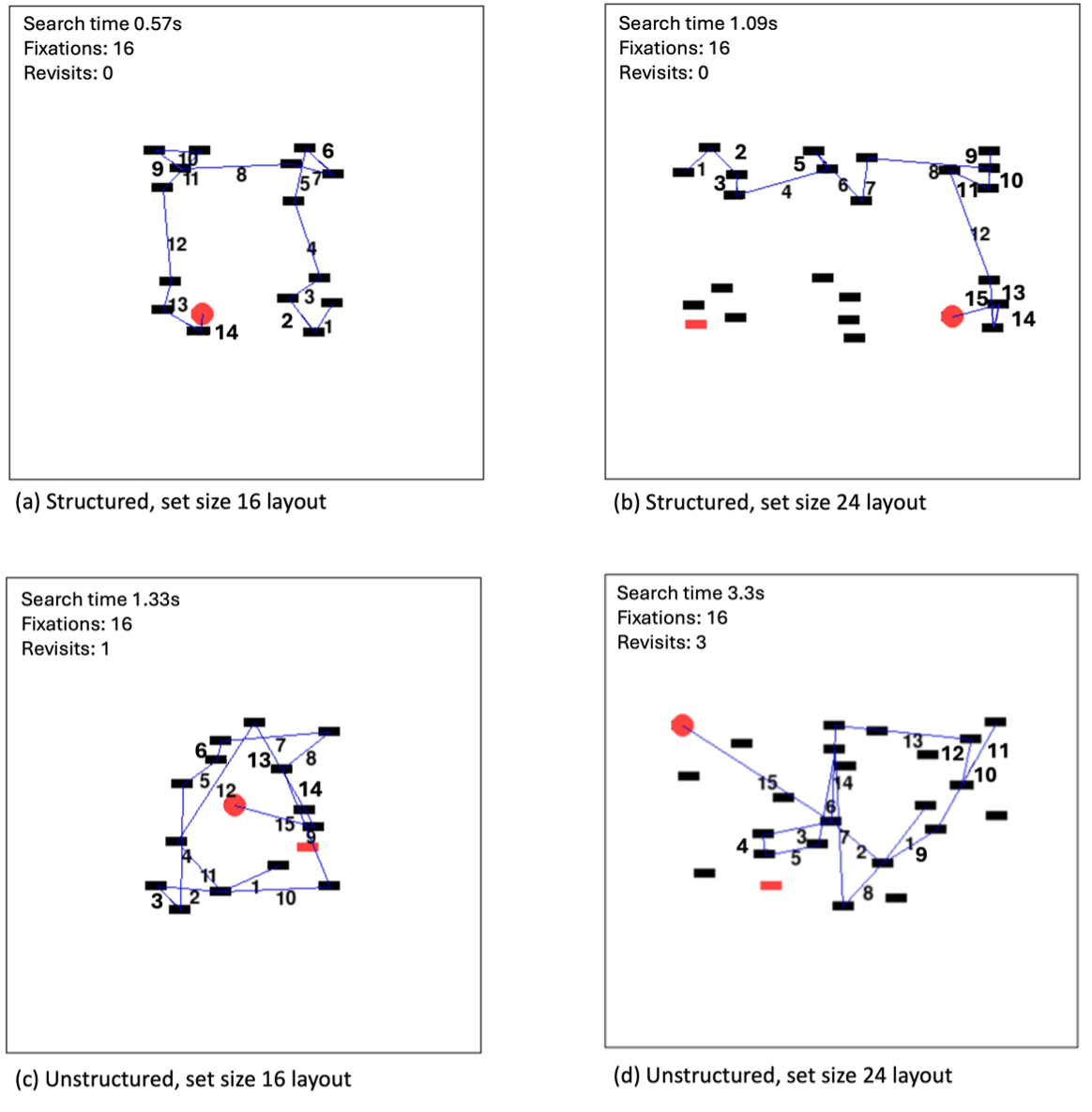}}
\caption{Examples of the eye movement simulations, comparing the number of fixations and revisits until the eye movement has progressed through 16 fixations from the beginning of the task. The numbering indicates saccades step by step, the red element represents the target element being searched for, and the red circle marks the eye's current position.\label{fig:fig_paths}}
\end{figure*}

In figure \ref{fig:fig_paths}, with a structured layout, the model's search strategy is optimized to search for targets one spatial group at a time. 
This shortens the length of the search path and search time and reduces the number of fixations by avoiding revisits to the same elements. 
The model cannot perceive structures as clearly in an unstructured layout. This leads to a situation where the model can scan through the elements to some extent, group by group, but not as efficiently as with a structured layout. 
Although our model seems to accurately predict the duration of visual search, we cannot confirm our assumption of human adaptation to a hierarchical search strategy based solely on our experiment. 

% TODO: discuss these limitations:
% 1. how is the hierarcical structure created?
% 2. why and how do the hierarcically different memory levels decay differently

\section{Discussion}

Our research contributes to advancing the understanding of how spatial structures influence the emergence of search strategies in visual search tasks. 
Cognitively rational adaptability has been demonstrated in various interactive tasks, providing explanatory power for how humans learn to harness environmental information, enabled by cognitive resources \cite{chen2015emergence, jokinen2020adaptive, jokinen2021multitasking, jokinen2021touchscreen, todi2019individualising}. 
Visually hierarchical structures enable an effective reduction of the search area in terms of memory. 
This behavior arises as a result of adaptation, when cognition has learned to harness the hierarchical capacity of visual memory to effectively reduce inhibition of return to already scanned elements. 
The adaptability of the model is a significant improvement over previous search models that predict hierarchical search, as the development of the search strategy emerges from the task goal, cognitive capacity, and environmental constraints, rather than being based on assumptions about the logic of the search strategy. 
Our model's search strategy is learned through reinforcement, utilizing indirect information from the environment. 
Additionally, the research reinforces earlier findings on the hierarchical nature of visual memory structure, showing how short-term flexible memory capacity enables memory expansion through chunking \cite{brady2011hierarchical, brady2013probabilistic, graham2000traveling, kong2010high}.

\subsection{Contribution}
Our model successfully predicts human eye movements in visual search tasks for both structured and unstructured layouts, as well as across a wide range of set sizes. 
The model adapted well to all layout variations without the need for parameter tuning. 
In order for our results to conclusively demonstrate optimization of eye movement strategies by encoding one structure at a time, our model requires further validation with human eye movement behavior. 

We believe that computational rational decision-making will enhance the model's accuracy across different conditions and explain their advantages for the search task, shaped by cognitive capacity and constraints. 
The model learns to utilize environmental information through hierarchical memory capacity to efficiently inhibit already encoded visual elements. 
The key difference between our model and previous models of hierarchical visual search is that our model’s hierarchical search strategy emerges as a result of adaptation, rather than being pre-programmed with logical rules. 
Previous models have encompassed the phenomenon of inhibition of return (IOR), but they have not yet been able to explain how a model learns to use hierarchical memory when adapting to visually hierarchical structures.

Our results add value to empirical research, as, contrary to the conclusions of Brumby and Zhuang \cite{brumby2015visual}, our findings indicate that visual grouping itself provides an advantage in search performance, even without semantic grouping. 
However, it is noteworthy that the stimuli in their experiment, in addition to line box grouping, also include spatial list grouping, which effectively introduces an additional hierarchical layer. 
It is therefore possible that too many hierarchical layers create extraneous cognitive load, meaning that the capacity of visual memory would be more optimal with fewer hierarchical layers. 
Since previous results have suggested that hierarchical structures guide visual search \cite{brumby2015visual, bailly2014model, chen2015emergence, halverson2008effects, halverson2011computational, hornof2001visual}, overly complex hierarchies might unnecessarily burden attention during information search. 
However, this requires further investigation to understand what limits visual hierarchies enable for the capacity to process information effectively, and when they may potentially have a negative effect on information search.

\subsection{Limitations}

Firstly, the real bottleneck of the model is the limited visual memory, which is rigid in creating chunked information in memory with structures of different sizes. 
This is possibly due to the computational complexity of the model in managing too many encoded chunks at once for the benefit of the search task. 
Compared to our model, the human visual system is likely far more dynamic, capable of perceiving structures during a search task and forming coherent information of all sizes in memory in a meaningful manner. 
The visual system is possibly able to expand structures from smaller structures and create new levels of hierarchical memory structure. 
Although the locations of individual elements can be encoded into short-term memory with a limited capacity of 4–6 items \cite{cowan2001magical, vogel2001storage, luck1997capacity}, it remains unresolved what the limitations of the hierarchical memory structure are.

Third significant weakness of the model is the limitation of visual memory's hierarchical structure to more than two levels. 
This explains partly, why the model increasingly diverges from participants in search times as the number of elements grows. 
The detailed analysis of eye movements revealed a weakness in the model's ability to learn optimally with larger numbers of elements, where its computational limits are reached. 
However, it is notable that the model's hierarchical VSTM contains only a limited two layers. 
It is more likely that human hierarchical visual memory is significantly more dynamic, capable of continuously creating new and multiple hierarchical layers from smaller structures.

A third significant limitation is our model's restricted ability to perceive visual structures. 
The model's symbolic representation of spatial grouping presupposes a rigid assumption regarding how spatial arrangement is perceived by the model. 
Application to UI design would require solving low-level processes of visual perception, particularly how visual structures are perceived directly in a pixel-based graphical interface. 
It would be possible to integrate promising clustering algorithms for visual grouping process.

\subsection{Conclusions}
At least partially tackling the aforementioned shortcomings in the model would enable the design of a real application as a tool to assist the designer's decision-making at the early stages of the design process. 
The model’s eye movement simulation also provides designers with educational insights into how cognitive limitations explain rational eye movement decision-making for a particular information structure in a layout. 
An essential developmental step for the model from the perspective of HCI research would be to also take semantic hierarchy into account with visual structures. 
This is a crucial design problem in the early stages of design: How to construct visual structures in a way that they are as consistent as possible with the semantic hierarchy \cite{brumby2015visual, niemela2003layout, salmeron2005expert}.

Our model demonstrates how the hierarchical visual capacity enables the utilization of visual structures as a result of adaptation. 
The flexible behavior of visual memory is a crucial area of research for gaining a deeper understanding of the role of cognitive limitations in information search. 
As we previously noted regarding the model's memory rigid limitations in relation to the presumed cognitive capacity of humans, a deeper understanding of the capacity of the hierarchical visual memory structure requires further research. 
To overcome the limitations of the model, it would be essential to determine what kind of flexible yet constrained visual memory enables efficient adaptation to environmental structures that contain large amounts of information.

%\FloatBarrier

\bibliographystyle{plain}

%\bibliographystyle{plain}
%\bibliography{references}

\begin{thebibliography}{10}

\bibitem{baddeley2003working}
Alan Baddeley.
\newblock Working memory: looking back and looking forward.
\newblock {\em Nature reviews neuroscience}, 4(10):829--839, 2003.

\bibitem{bailly2014model}
Gilles Bailly, Antti Oulasvirta, Duncan~P Brumby, and Andrew Howes.
\newblock Model of visual search and selection time in linear menus.
\newblock In {\em Proceedings of the sigchi conference on human factors in
  computing systems}, pages 3865--3874, 2014.

\bibitem{bays2009precision}
Paul~M Bays, Raquel~FG Catalao, and Masud Husain.
\newblock The precision of visual working memory is set by allocation of a
  shared resource.
\newblock {\em Journal of vision}, 9(10):7--7, 2009.

\bibitem{brady2011hierarchical}
Timothy~F Brady and George~A Alvarez.
\newblock Hierarchical encoding in visual working memory: Ensemble statistics
  bias memory for individual items.
\newblock {\em Psychological science}, 22(3):384--392, 2011.

\bibitem{brady2009compression}
Timothy~F Brady, Talia Konkle, and George~A Alvarez.
\newblock Compression in visual working memory: using statistical regularities
  to form more efficient memory representations.
\newblock {\em Journal of Experimental Psychology: General}, 138(4):487, 2009.

\bibitem{brady2013probabilistic}
Timothy~F Brady and Joshua~B Tenenbaum.
\newblock A probabilistic model of visual working memory: Incorporating higher
  order regularities into working memory capacity estimates.
\newblock {\em Psychological review}, 120(1):85, 2013.

\bibitem{brumby2015visual}
Duncan~P Brumby and Susan Zhuang.
\newblock Visual grouping in menu interfaces.
\newblock In {\em Proceedings of the 33rd Annual ACM Conference on Human
  Factors in Computing Systems}, pages 4203--4206, 2015.

\bibitem{bundesen1983color}
Claus Bundesen and Leif~Flemming Pedersen.
\newblock Color segregation and visual search.
\newblock {\em Perception \& Psychophysics}, 33:487--493, 1983.

\bibitem{chase1973mind}
William~G Chase and Herbert~A Simon.
\newblock The mind's eye in chess.
\newblock In {\em Visual information processing}, pages 215--281. Elsevier,
  1973.

\bibitem{chen2015emergence}
Xiuli Chen, Gilles Bailly, Duncan~P Brumby, Antti Oulasvirta, and Andrew Howes.
\newblock The emergence of interactive behavior: A model of rational menu
  search.
\newblock In {\em Proceedings of the 33rd annual ACM conference on human
  factors in computing systems}, pages 4217--4226, 2015.

\bibitem{cowan2001magical}
Nelson Cowan.
\newblock The magical number 4 in short-term memory: A reconsideration of
  mental storage capacity.
\newblock {\em Behavioral and brain sciences}, 24(1):87--114, 2001.

\bibitem{deco2007attention}
Gustavo Deco and Dietmar Heinke.
\newblock Attention and spatial resolution: A theoretical and experimental
  study of visual search in hierarchical patterns.
\newblock {\em Perception}, 36(3):335--354, 2007.

\bibitem{findlay1998eye}
John~M Findlay and Iain~D Gilchrist.
\newblock Eye guidance and visual search.
\newblock In {\em Eye guidance in reading and scene perception}, pages
  295--312. Elsevier, 1998.

\bibitem{gershman2015computational}
Samuel~J Gershman, Eric~J Horvitz, and Joshua~B Tenenbaum.
\newblock Computational rationality: A converging paradigm for intelligence in
  brains, minds, and machines.
\newblock {\em Science}, 349(6245):273--278, 2015.

\bibitem{graham2000traveling}
Scott~M Graham, Anupam Joshi, and Zygmunt Pizlo.
\newblock The traveling salesman problem: A hierarchical model.
\newblock {\em Memory \& cognition}, 28:1191--1204, 2000.

\bibitem{halverson2008effects}
Tim Halverson and Anthony~J Hornof.
\newblock The effects of semantic grouping on visual search.
\newblock In {\em CHI'08 Extended Abstracts on Human Factors in Computing
  Systems}, pages 3471--3476. 2008.

\bibitem{halverson2011computational}
Tim Halverson and Anthony~J Hornof.
\newblock A computational model of “active vision” for visual search in
  human--computer interaction.
\newblock {\em Human--Computer Interaction}, 26(4):285--314, 2011.

\bibitem{ho2022people}
Mark~K Ho, David Abel, Carlos~G Correa, Michael~L Littman, Jonathan~D Cohen,
  and Thomas~L Griffiths.
\newblock People construct simplified mental representations to plan.
\newblock {\em Nature}, 606(7912):129--136, 2022.

\bibitem{hochstein2002view}
Shaul Hochstein and Merav Ahissar.
\newblock View from the top: Hierarchies and reverse hierarchies in the visual
  system.
\newblock {\em Neuron}, 36(5):791--804, 2002.

\bibitem{hornof2001visual}
Anthony~J Hornof.
\newblock Visual search and mouse-pointing in labeled versus unlabeled
  two-dimensional visual hierarchies.
\newblock {\em ACM Transactions on Computer-Human Interaction (TOCHI)},
  8(3):171--197, 2001.

\bibitem{humphreys1993search}
Glyn~W Humphreys and Hermann~J Muller.
\newblock Search via recursive rejection (serr): A connectionist model of
  visual search.
\newblock {\em Cognitive Psychology}, 25(1):43--110, 1993.

\bibitem{itti2000saliency}
Laurent Itti and Christof Koch.
\newblock A saliency-based search mechanism for overt and covert shifts of
  visual attention.
\newblock {\em Vision research}, 40(10-12):1489--1506, 2000.

\bibitem{itti2001feature}
Laurent Itti and Christof Koch.
\newblock Feature combination strategies for saliency-based visual attention
  systems.
\newblock {\em Journal of Electronic imaging}, 10(1):161--169, 2001.

\bibitem{jiang2004kind}
Yuhong Jiang and Stephanie~W Wang.
\newblock What kind of memory supports visual marking?
\newblock {\em Journal of Experimental Psychology: Human Perception and
  Performance}, 30(1):79, 2004.

\bibitem{jokinen2021touchscreen}
Jussi Jokinen, Aditya Acharya, Mohammad Uzair, Xinhui Jiang, and Antti
  Oulasvirta.
\newblock Touchscreen typing as optimal supervisory control.
\newblock In {\em Proceedings of the 2021 CHI conference on human factors in
  computing systems}, pages 1--14, 2021.

\bibitem{jokinen2021multitasking}
Jussi~PP Jokinen, Tuomo Kujala, and Antti Oulasvirta.
\newblock Multitasking in driving as optimal adaptation under uncertainty.
\newblock {\em Human factors}, 63(8):1324--1341, 2021.

\bibitem{jokinen2020adaptive}
Jussi~PP Jokinen, Zhenxin Wang, Sayan Sarcar, Antti Oulasvirta, and Xiangshi
  Ren.
\newblock Adaptive feature guidance: Modelling visual search with graphical
  layouts.
\newblock {\em International Journal of Human-Computer Studies}, 136:102376,
  2020.

\bibitem{kahneman1992reviewing}
Daniel Kahneman, Anne Treisman, and Brian~J Gibbs.
\newblock The reviewing of object files: Object-specific integration of
  information.
\newblock {\em Cognitive psychology}, 24(2):175--219, 1992.

\bibitem{klein1999inhibition}
Raymond~M Klein and W~Joseph MacInnes.
\newblock Inhibition of return is a foraging facilitator in visual search.
\newblock {\em Psychological science}, 10(4):346--352, 1999.

\bibitem{kong2010high}
Xiaohui Kong, Christian~D Schunn, and Garrick~L Wallstrom.
\newblock High regularities in eye-movement patterns reveal the dynamics of the
  visual working memory allocation mechanism.
\newblock {\em Cognitive science}, 34(2):322--337, 2010.

\bibitem{li2018memory}
Chia-Ling Li, M~Pilar Aivar, Matthew~H Tong, and Mary~M Hayhoe.
\newblock Memory shapes visual search strategies in large-scale environments.
\newblock {\em Scientific reports}, 8(1):4324, 2018.

\bibitem{luck1997capacity}
Steven~J Luck and Edward~K Vogel.
\newblock The capacity of visual working memory for features and conjunctions.
\newblock {\em Nature}, 390(6657):279--281, 1997.

\bibitem{ma2014changing}
Wei~Ji Ma, Masud Husain, and Paul~M Bays.
\newblock Changing concepts of working memory.
\newblock {\em Nature neuroscience}, 17(3):347--356, 2014.

\bibitem{miller1956magical}
George~A Miller.
\newblock The magical number seven, plus or minus two: Some limits on our
  capacity for processing information.
\newblock {\em Psychological review}, 63(2):81, 1956.

\bibitem{mitchell2008flexible}
Daniel~J Mitchell and Rhodri Cusack.
\newblock Flexible, capacity-limited activity of posterior parietal cortex in
  perceptual as well as visual short-term memory tasks.
\newblock {\em Cerebral cortex}, 18(8):1788--1798, 2008.

\bibitem{najemnik2005optimal}
Jiri Najemnik and Wilson~S Geisler.
\newblock Optimal eye movement strategies in visual search.
\newblock {\em Nature}, 434(7031):387--391, 2005.

\bibitem{nassar2018chunking}
Matthew~R Nassar, Julie~C Helmers, and Michael~J Frank.
\newblock Chunking as a rational strategy for lossy data compression in visual
  working memory.
\newblock {\em Psychological review}, 125(4):486, 2018.

\bibitem{navalpakkam2005modeling}
Vidhya Navalpakkam and Laurent Itti.
\newblock Modeling the influence of task on attention.
\newblock {\em Vision research}, 45(2):205--231, 2005.

\bibitem{niemela2003layout}
Marketta Niemel{\"a} and Pertti Saariluoma.
\newblock Layout attributes and recall.
\newblock {\em Behaviour \& information technology}, 22(5):353--363, 2003.

\bibitem{nikolic2007creation}
Danko Nikoli{\'c} and Wolf Singer.
\newblock Creation of visual long-term memory.
\newblock {\em Perception \& psychophysics}, 69(6):904--912, 2007.

\bibitem{oh2004role}
Sei-Hwan Oh and Min-Shik Kim.
\newblock The role of spatial working memory in visual search efficiency.
\newblock {\em Psychonomic bulletin \& review}, 11:275--281, 2004.

\bibitem{osborne2010improving}
Jason Osborne.
\newblock Improving your data transformations: Applying the box-cox
  transformation.
\newblock {\em Practical Assessment, Research, and Evaluation}, 15(1), 2010.

\bibitem{oulasvirta2022computational}
Antti Oulasvirta, Jussi~PP Jokinen, and Andrew Howes.
\newblock Computational rationality as a theory of interaction.
\newblock In {\em Proceedings of the 2022 CHI Conference on Human Factors in
  Computing Systems}, pages 1--14, 2022.

\bibitem{parkhurst2002modeling}
Derrick Parkhurst, Klinton Law, and Ernst Niebur.
\newblock Modeling the role of salience in the allocation of overt visual
  attention.
\newblock {\em Vision research}, 42(1):107--123, 2002.

\bibitem{pomerantz1973role}
James~R Pomerantz and WR~Garner.
\newblock The role of configuration and target discriminability in a visual
  search task.
\newblock {\em Memory \& Cognition}, 1(1):64--68, 1973.

\bibitem{pomplun2003area}
Marc Pomplun, Eyal~M Reingold, and Jiye Shen.
\newblock Area activation: A computational model of saccadic selectivity in
  visual search.
\newblock {\em Cognitive Science}, 27(2):299--312, 2003.

\bibitem{posner1984components}
Michael~I Posner, Yoav Cohen, et~al.
\newblock Components of visual orienting.
\newblock {\em Attention and performance X: Control of language processes},
  32:531--556, 1984.

\bibitem{posner1985inhibition}
Michael~I Posner, Robert~D Rafal, Lisa~S Choate, and Jonathan Vaughan.
\newblock Inhibition of return: Neural basis and function.
\newblock {\em Cognitive neuropsychology}, 2(3):211--228, 1985.

\bibitem{rao2002eye}
Rajesh~PN Rao, Gregory~J Zelinsky, Mary~M Hayhoe, and Dana~H Ballard.
\newblock Eye movements in iconic visual search.
\newblock {\em Vision research}, 42(11):1447--1463, 2002.

\bibitem{rayner1978eye}
Keith Rayner.
\newblock Eye movements in reading and information processing.
\newblock {\em Psychological bulletin}, 85(3):618, 1978.

\bibitem{salmeron2005expert}
Ladislao Salmer{\'o}n, Jos{\'e}~J Ca{\~n}as, and Inmaculada Fajardo.
\newblock Are expert users always better searchers? interaction of expertise
  and semantic grouping in hypertext search tasks.
\newblock {\em Behaviour \& information technology}, 24(6):471--475, 2005.

\bibitem{salvucci2001integrated}
Dario~D Salvucci.
\newblock An integrated model of eye movements and visual encoding.
\newblock {\em Cognitive Systems Research}, 1(4):201--220, 2001.

\bibitem{snyder2000inhibition}
Janice~J Snyder and Alan Kingstone.
\newblock Inhibition of return and visual search: How many separate loci are
  inhibited?
\newblock {\em Perception \& Psychophysics}, 62(3):452--458, 2000.

\bibitem{sun2008computer}
Yaoru Sun, Robert Fisher, Fang Wang, and Herman~Martins Gomes.
\newblock A computer vision model for visual-object-based attention and eye
  movements.
\newblock {\em Computer vision and image understanding}, 112(2):126--142, 2008.

\bibitem{teo2012cogtool}
Leong-Hwee Teo, Bonnie John, and Marilyn Blackmon.
\newblock Cogtool-explorer: A model of goal-directed user exploration that
  considers information layout.
\newblock In {\em Proceedings of the SIGCHI conference on human factors in
  computing systems}, pages 2479--2488, 2012.

\bibitem{todi2019individualising}
Kashyap Todi, Jussi Jokinen, Kris Luyten, and Antti Oulasvirta.
\newblock Individualising graphical layouts with predictive visual search
  models.
\newblock {\em ACM Transactions on Interactive Intelligent Systems (TiiS)},
  10(1):1--24, 2019.

\bibitem{treisman1982perceptual}
Anne Treisman.
\newblock Perceptual grouping and attention in visual search for features and
  for objects.
\newblock {\em Journal of experimental psychology: human perception and
  performance}, 8(2):194, 1982.

\bibitem{treisman1980feature}
Anne~M Treisman and Garry Gelade.
\newblock A feature-integration theory of attention.
\newblock {\em Cognitive psychology}, 12(1):97--136, 1980.

\bibitem{van2020embodied}
Stefan Van~der Stigchel.
\newblock An embodied account of visual working memory.
\newblock {\em Visual cognition}, 28(5-8):414--419, 2020.

\bibitem{vogel2001storage}
Edward~K Vogel, Geoffrey~F Woodman, and Steven~J Luck.
\newblock Storage of features, conjunctions, and objects in visual working
  memory.
\newblock {\em Journal of experimental psychology: human perception and
  performance}, 27(1):92, 2001.

\bibitem{wang2010searching}
Zhiguo Wang and Raymond~M Klein.
\newblock Searching for inhibition of return in visual search: A review.
\newblock {\em Vision research}, 50(2):220--228, 2010.

\bibitem{watson1997visual}
Derrick~G Watson and Glyn~W Humphreys.
\newblock Visual marking: prioritizing selection for new objects by top-down
  attentional inhibition of old objects.
\newblock {\em Psychological review}, 104(1):90, 1997.

\bibitem{wolfe1994guided}
Jeremy~M Wolfe.
\newblock Guided search 2.0 a revised model of visual search.
\newblock {\em Psychonomic bulletin \& review}, 1:202--238, 1994.

\bibitem{wolfe1994visual}
Jeremy~M Wolfe.
\newblock Visual search in continuous, naturalistic stimuli.
\newblock {\em Vision research}, 34(9):1187--1195, 1994.

\bibitem{woodman2003perceptual}
Geoffrey~F Woodman, Shaun~P Vecera, and Steven~J Luck.
\newblock Perceptual organization influences visual working memory.
\newblock {\em Psychonomic bulletin \& review}, 10(1):80--87, 2003.

\end{thebibliography}

\appendix
\section{The Generation of the Layout}
At the beginning of each new search task, a layout was generated for the model using an algorithm, which was also used for generating layouts for the experiment. Layouts were generated for structured and unstructured conditions and both conditions for four different element sizes (16, 24, 36, 48). 
The elements were partially randomized for the layout with certain constraints. 
In a structured layout, the elements are arranged spatially in groups of four elements, but the positions of the elements are randomized within the area of the group, and the minimum distance between elements is 1.7 times the width of the element. 
In the unstructured layout, elements are otherwise arranged randomly, but the minimum distance between elements is at least 2.6 times the width of an element. 
In generation algorithm, a random location for the element is drawn as many times as necessary until the minimum distance criterion is met in relation to other elements. 
The minimum distance is larger for the unstructured layout because it ensures that mere randomization of location does not produce too obvious structures.

%Layouts in Figure illustrates how, according to the generation algorithm, elements are placed in different areas and grouped into the model's internal representation. 
In generating the layout, the structure areas are first defined, with their centroids spaced as evenly as possible from each other. 
Subsequently, the size of the area is defined, which for a structured layout is small enough so that the elements are sufficiently close to each other within the structure (spatially grouped). 
For the unstructured layout, the structure areas are significantly larger, resulting in areas partially overlapping each other. 
%According to tables of the Figure, this is reflected in the model's internal representation in such a way that, in the case of the structured layout, each element can only belong to one group. 
In the case of unstructured layouts, elements primarily belong to one group, but some elements may belong to more than one group in the internal representation because their location may also belong to the area of another group. 
To ensure that spatial proximity itself does not cause faster search times on a structured layout, spatial structures are arranged so far apart from each other that, at the same set size, the average distance between elements is greater than on an unstructured layout.

\end{document}